\documentclass[conference]{IEEEtran}
\IEEEoverridecommandlockouts
\usepackage{enumitem}
\usepackage{colortbl}
\definecolor{lightgray}{gray}{0.9}
 \pdfoutput=1 
\usepackage{hyperref} 

\usepackage{cite}
\usepackage{amsmath,amssymb,amsfonts}
\usepackage{algorithmic}
\usepackage{graphicx}
\usepackage{textcomp}
\usepackage{xcolor}
\def\BibTeX{{\rm B\kern-.05em{\sc i\kern-.025em b}\kern-.08em
    T\kern-.1667em\lower.7ex\hbox{E}\kern-.125emX}}
\begin{document}

\title{Fine-Tuning LLMs for Code Mutation: A New Era of Cyber Threats}

\author{\IEEEauthorblockN{Mohammad Setak}
\IEEEauthorblockA{
\textit{Computer Science} \\
\textit{Ontario Tech University}\\
Oshawa, Canada \\
mohammadhossein.setak@ontariotechu.net}
\and
\IEEEauthorblockN{Pooria Madani}
\IEEEauthorblockA{\textit{Business and Information Technology} \\
\textit{Ontario Tech University}\\
Oshawa, Canada \\
pooria.madani@ontariotechu.ca}
\and
}

\maketitle

\begin{abstract}
Recent advancements in Large Language Models (LLMs) have significantly improved their capabilities in natural language processing and code synthesis, enabling more complex applications across different fields. This paper explores the application of LLMs in the context of code mutation, a process where the structure of program code is altered without changing its functionality. Traditionally, code mutation has been employed to increase software robustness in mission-critical applications. Additionally, mutation engines have been exploited by malware developers to evade the signature-based detection methods employed by malware detection systems. Existing code mutation engines, often used by such threat actors, typically result in only limited variations in the malware, which can still be identified through static code analysis. However, the agility demonstrated by an LLM-based code synthesizer could significantly change this threat landscape by allowing for more complex code mutations that are not easily detected using static analysis. One can increase variations of codes synthesized by a pre-trained LLM through fine-tuning and retraining. This process is what we refer to as code mutation training. In this paper, we propose a novel definition of code mutation training tailored for pre-trained LLM-based code synthesizers and demonstrate this training on a lightweight pre-trained model. Our approach involves restructuring (i.e., mutating) code at the subroutine level, which allows for more manageable mutations while maintaining the semantic integrity verified through unit testing. Our experimental results illustrate the effectiveness of our approach in improving code mutation capabilities of LLM-based program synthesizers in producing varied and functionally correct code solutions, showcasing their potential to transform the landscape of code mutation and the threats associated with it.
\end{abstract}


\section{Introduction}
Code mutation refers to the process of altering a computer program's source code in such a way that, while the core behaviour of the program remains unchanged, its structure or representation is modified. Any modification that alters the syntax of a program without changing its semantics is considered a mutation of the source code. As a pivotal aspect of software engineering, code mutation can be used for various purposes including testing the reliability of software, improving security measures against malware detection, and developing more resilient systems against cyber-attacks \cite{madani2023metamorphic}.

However, the utility of code mutation extends into various (malicious) domains as well, such as the development of metamorphic malware \cite{you2010malware}. Metamorphic malware is designed to alter its codebase each time it replicates itself. It preserves its core functionality while modifying its code structure \cite{sun2016finding}. This allows the malware to evade signature-based malware detection systems.
Traditionally, metamorphic malware has depended on rule-based mutation engines. While effective to some extent, these engines operate within a constrained set of predefined rules for modifying a given codebase, such as reordering instructions, substituting synonymous operations, or inserting non-functional code (i.e., NOP instruction). However, the mutations generated by these rule-based systems tend to be predictable and can often be identified through static code analysis techniques \cite{zhang2007metaaware}.

Large Language Models (LLMs) are generative learning-based models extensively trained on a broad spectrum of textual data (e.g., poems, news articles, and program source codes) \cite{chen2021evaluating}. This varied and extensive training enables models to capture distance dependencies in a sequence. Therefore, these models can be very effective in sequence generative tasks such as those in NLP (e.g., sentiment analysis, language translation, etc.) and also program code synthesis \cite{ghaemi2024transformers}. In addition to traditional text sources, the training data for these models often includes vast amounts of code from various online repositories and forums where developers share and discuss programming challenges \cite{zhao2023survey}. By incorporating these coding resources into their training datasets, LLMs have demonstrated the ability to synthesize program code. This capability allows them to assist in software development tasks such as code completion, error correction, and even generating entire scripts or applications from scratch based on high-level descriptions \cite{feng2020codebert}.

Unlike rule-based systems, LLMs can generate a vast array of novel code variations based on a deep semantic understanding of a given code, potentially evading the capabilities of static analysis tools to detect or predict code mutation \cite{madani2023metamorphic}. This advancement not only challenges existing cybersecurity defenses but also demonstrates how intelligent metamorphic malware can regenerate malwares previously detected. As LLMs vary in their ability to synthesize code, they demonstrate differing efficiencies in the domain of code mutation as well. Effective code mutation requires the ability to not just solve programming problems but to solve them in multiple ways, each with unique syntactical variations. 

Code synthesizing and code mutation represent two distinct capabilities of LLMs. Code synthesizing focuses on synthesizing correct and efficient code from high-level descriptions and is a direct measure of an LLM’s capability to execute programming tasks. On the other hand, code mutation assesses an LLM’s capacity to reframe or restructure existing code in numerous functionally equivalent but syntactically different ways.  While one LLM-based model might excel in generating solutions across a broad range of problems with similar solutions for each one, another might solve a narrower array of problems but with more variation in each solution's structure and syntax.
As the code synthesizing ability of an LLM can be enhanced through retraining and fine-tuning, the same techniques can be applied to improve its code mutation capabilities. This paper proposes a structured approach to improve pre-trained LLM models specifically for code mutation tasks. 

In this paper, we propose a definition for code mutation training at the subroutine level using LLM-based code synthesizers. Since programs are typically composed of subroutines, performing code mutations at the subroutine level equates to mutating the entire program. Therefore, to facilitate code mutation, the task is segmented into smaller, more manageable sub-tasks of mutating individual subroutines. Moreover, since subroutines are well-defined and responsible for simpler tasks than the entire program, mutating their code becomes a much more feasible task for an LLM-based code synthesizer as a mutation engine. Additionally, with each subroutine accompanied by a unit test, verifying the semantic integrity of these mutated codes becomes straightforward. This approach ensures that despite changes in syntax, the functionality remains consistent and accurate.

A few state-of-the-art pre-trained LLMs, such as Llama3 by Meta \cite{dubey2024llama}, have demonstrated great code mutability capabilities. However, due to their disk sizes, embedding these code synthesizers in a (malicious) software for just-in-time code mutation is impractical. Threat actors are actively in search of smaller and more compact LLM-based code synthesis/mutating engines in order to be embedded in their malware. In this work, specifically, we attempt to study this objective systematically. Our contributions are threefold as follows:

\begin{enumerate}
    \item Propose a formal definition for a learning-based code mutation task,
    \item Employ a novel iterative framework to generate (a publicly available) \footnote{\url{https://github.com/mhstk/code-mutation-dataset}} code mutation training dataset using a teacher LLM,
    \item Demonstrate code mutation training on a lightweight code synthesizer LLM and evaluate its effectiveness.
\end{enumerate}

\medskip
Our work clearly demonstrates the success of training a (relatively) lightweight LLM, showcasing its potential in code mutation. This advancement could influence the development of metamorphic malware, underscoring the need for next-generation malware detection systems to be equipped with mechanisms that can effectively detect these evolving threats.

The rest of the paper is organized as follows: In section 2 we have provided a detailed overview of existing works in code synthesizer LLMs. Section 3 provides our definition of code mutation training and how it is evaluated. In section 4, we address the dataset challenge for code mutation training and propose a framework to create a training dataset for this purpose. Section 5 demonstrates code mutation training in LLMs through experiments, examines the results, and discusses the findings. Finally, in Section 6 we conclude the work and outline potential future works.

\bigskip

\section{Related Works}
The utility of machine learning models in understanding and replicating the distribution of their training data is foundational to their ability to perform tasks ranging from simple classification to complex generative processes. Generative models, a subset of machine learning models, are particularly designed to learn the joint probability distribution of features within a given training dataset and utilize this learning to generate new data instances that are statistically similar to the original data points used in training \cite{theis2015note}.

Generative models play an important role in many NLP tasks, such as text summarization, sentiment analysis, and text translation. LLMs are advanced generative models that have redefined performance benchmarks in these areas. The recent significant improvements in NLP tasks have been driven by the advent of LLMs, which are characterized by their deep neural network architectures and substantial trainable parameter sizes \cite{du2020novel}. These models are trained on massive textual datasets to generate outputs that can mimic human-like understanding and coherence \cite{zhao2023survey}.

The significant impact of LLMs began with the introduction of the Transformer architecture, as described by Vaswani \textit{et al}. \cite{vaswani2017attention} in their seminal paper, ``Attention is All You Need". This architecture departs from prior sequence-based modeling techniques by employing mechanisms like self-attention and positional encoding to process sequences in parallel, significantly improving efficiency and scalability in handling long sequences. This architectural advancement laid the groundwork for several foundational models in NLP, including BERT (Bidirectional Encoder Representations from Transformers) \cite{devlin2018bert} and GPT(Generative Pre-trained Transformer) \cite{brown2020language}. Developed by Google, BERT analyzes text by simultaneously processing a word sequence in both directions, demonstrating great efficacy in understanding the context of words within text sequences. On the other hand, GPT, developed by OpenAI, processes input sequences from left to right and is trained to predict the next element in the sequence using an auto-regressive approach.

As LLMs have proven effective in generating text sequences in natural language, they are also shown to similarly excel in synthesizing software code (i.e., code or program synthesis) written in different programming languages \cite{feng2020codebert}. Program synthesis using LLMs involves generating executable code based on high-level specifications or descriptions provided in natural language. In this section, we will review some of the existing works in computer program synthesis using transformers and pre-trained large language models.

CodeBERT by Feng \textit{et al}. \cite{feng2020codebert} is a bimodal BERT model that blends the boundaries between natural and programming languages. It is specifically designed to understand and generate code by being pre-trained on a dataset consisting of both natural language and code. This model has shown effectiveness in tasks like code documentation, which involves generating natural language descriptions from code snippets, and vice versa.

GraphCodeBERT \cite{guo2020graphcodebert} introduces two innovative structure-aware pre-training tasks specifically designed to learn representations from source code. These tasks utilize data flow graphs that represent variable relationships within the Abstract Syntax Tree (AST) format.

PLBART \cite{ahmad2021unified} is a sequence-to-sequence model that leverages the BART \cite{lewis2019bart} architecture. BART utilizes a transformer-based architecture, trained through a process where text is corrupted by an arbitrary noising function, followed by the model learning to reconstruct the original text. PLBART extends this approach by being pre-trained on a vast corpus of Java and Python functions along with associated natural language text, employing denoising autoencoding techniques.

OpenAI's Codex \cite{chen2021evaluating} builds on the capabilities of GPT model, fine-tuned to understand and generate code across several programming languages from natural language descriptions. The Codex is trained on a vast repository of GitHub codes for problem-solving. Codex is used to build famous applications such as ChatGPT and GitHub Copilot, which further demonstrate its utility in real-world programming environments by assisting developers with code completion and other programming tasks.

CodeGen, by Salesforce \cite{nijkamp2022codegen},  was trained sequentially across multiple datasets. Initially, the model was trained on ThePile dataset \cite{xu2022systematic}, a large and diverse dataset consisting of both code and text data, designed to foster a comprehensive understanding of human language. Training then progressed to BigQuery dataset \cite{hoffa2016github}, which integrates GitHub data with Google BigQuery to enhance the model's capabilities in handling complex data structures and queries. Finally, the training concluded with BigPython, a dataset that focuses exclusively on Python code, aiding in refining the model’s proficiency in coding syntax and logic.

AlphaCode, developed by Li \textit{et al.} at Google \cite{li2022competition}, takes the problem a step further by not only generating code snippets but also whole programs that can compete at a professional level in coding competitions. This model was trained using a diverse set of competitive programming problems and represents a substantial leap towards automating complex problem-solving through code.

Code Llama \cite{roziere2023code} is a family of large language models tailored for coding tasks, leveraging the Llama 2 architecture. It achieves state-of-the-art performance among open models and features notable capabilities such as infilling, support for large input contexts, and zero-shot instruction following for programming tasks. To accommodate a diverse range of applications, Code Llama offers several variants: foundational models (Code Llama), Python specializations (Code Llama - Python), and instruction-following models (Code Llama - Instruct). Each variant is available in multiple sizes, specifically designed to meet distinct computational and functional needs.

Llama3 \cite{dubey2024llama} by Meta represents a family of large language models, available in pre-trained and instruction-tuned generative versions. Accessible in both 8B and 70B parameter sizes, these models are particularly optimized for dialogue use cases. They outperform many available open-source chat models on common industry benchmarks. Additionally, Llama 3 excels in code synthesis, with its capabilities highlighted by impressive results on code generation benchmarks.

CodeT5 \cite{wang2021codet5} is a unified pre-trained encoder-decoder Transformer model that effectively leverages code semantics captured from developer-assigned identifiers. This model operates within a unified framework, facilitating seamless support for both code understanding and generation tasks. It also enables multi-task learning. A distinctive feature of CodeT5 is its novel identifier-aware pre-training task, which allows the model to recognize code tokens as identifiers and accurately recover them when masked.

CodeRL \cite{le2022coderl} introduces a novel framework for program synthesis tasks, combining pre-trained LLMs with deep reinforcement learning (RL). During training, CodeRL utilizes an actor network for code generation and a critic network that assesses the functional correctness of the generated programs. This critic network is specifically trained to provide dense feedback signals to the actor, thereby enhancing the learning process and improving the quality of the code output.

Table 1 presents an evaluation of some of the code synthesizer LLMs \cite{madani2023metamorphic, liu2024your}.

\begin{table}[htbp]
\caption{Comparing Code Synthesizer LLM models \\
with \textit{pass@k} for k=\{1, 10\} on the HumanEval dataset} 
\begin{center}
\begin{tabular}{|c|c|c|c|}
\hline
\textbf{Model} & \textbf{params} & \textbf{\textit{pass@1}} & \textbf{\textit{pass@10}} \\
\hline
\multicolumn{4}{|c|}{\cellcolor{lightgray}Closed source} \\
\hline
Codex \cite{chen2021evaluating} & 2.5B & 21.4 & 35.4 \\
\hline
GPT 3.5 \cite{radford2018improving} & 175B & 48.1 & - \\
\hline
GPT-4 \cite{achiam2023gpt} & 175B & 88.4 & - \\
\hline
AlphaCode \cite{li2022competition} & 1.1B & 17.1 & 28.2 \\
\hline
Phi-l \cite{gunasekar2023textbooks} & 1.3B & 50.6 & - \\
\hline
\multicolumn{4}{|c|}{\cellcolor{lightgray}Open source} \\
\hline
Codegen \cite{chen2021evaluating} & 6B & 27.7 & 46.9 \\
\hline
Codegen2 \cite{nijkamp2023codegen2} & 7B & 17.9 & 30.9 \\
\hline
CodeT5+ \cite{wang2023codet5+} & 2B & 24.2 & 38.2 \\
\hline
StarCoder \cite{li2023starcoder} & 15B & 32.2 & 56.7 \\
\hline
SantaCoder \cite{allal2023santacoder} & 1.1B & 16.6 & 29.2 \\
\hline
CodeLlama \cite{roziere2023code} & 7B & 37.8 & 39.2 \\
\hline
Llama3 \cite{dubey2024llama} & 8B & 62.2 & - \\
\hline
InCoder \cite{fried2022incoder} & 1.3B & 10 & 15.9 \\
\hline
PolyCoder \cite{xu2022systematic} & 2.7B & 5.9 & 10.2 \\
\hline
\end{tabular}
\label{tatheis2015note}
\end{center}
\end{table}

\section{Methodology}
Let $M^\theta$ denote a pre-trained program-synthesizer LLM with a learned parameter set $\theta$; let $P=\{p_1,p_2,\ldots,p_n\}$ denote the list of prompts to $M$ used for model validation such that each $p_i \in P$ starts with a subroutine (i.e., method/function) definition followed by multiple lines of human-readable comments articulating the semantics of the subroutine with no code implementation. Let 

\begin{equation}
    Synth:M^{\theta} \left(p_i\in P,k \right) \rightarrow S_i=\{s_i^j\}_{j=1}^k
\end{equation}
denotes a subroutine synthesizer by applying a prompt from $P$ to $M$ and sampling the output layer of the model $k$ times to generate a set of synthesized solution $S_i$ to the given prompt. Let $U=\{u_1,u_2,\ldots,u_n\}\ $ denote a set of unit tests where $u_i\in U$ is the corresponding unit test for output synthesized $S_i=\{Synth\left(p_i\in P,k\right)\} $\footnote{In this context, despite using the term `set' and its notation, we are permitting duplicate values within these sets.}. In an ideal setting, all synthesized outputs in $S_i$ are syntactically and semantically correct implementations of the prompt input $p_i$. Moreover, in an ideal setting, and as far as code mutation capabilities of $M^\theta$ is concerned, all synthesized solutions instances in $S_i$ pass the associated unit test $u_i$ while their syntactic structures are unique.
 
However, code synthesis and mutation capabilities of a given pre-trained $M^\theta$ LLM are far from ideal. In other words, for a given $p_i\in P$, not all synthesized solutions in $S_i$ are correct (i.e., do not pass the associated unit test $u_i$) or unique. Therefore synthesized $S_i$ could have repeated members as the model might produce identical codes. In the context of this work, two generated solutions are considered identical when they are syntactically identical. This equivalence can be established through character-by-character matching of the codes, identical abstract syntax trees, or any other method that represents syntactic identity.

We use the $pass@k$ metric proposed by  Kulal \textit{et al}. \cite{kulal2019spoc} to measure $M^{\theta}$’s ability to synthesize a correct coding solution to a given prompt $p_i$  For each problem $p_i\in P$, let $C_i$ denote a set that consists of all correct solutions from $s_i^j \in S_i$, that pass their respective unit test. Then, $pass@k$ is defined as a metric that assesses an LLM's ability to synthesize computer program code by calculating the fraction of problems solved relative to the total number of problems. A problem $p_i$ is considered solved if there is at least one solution $s_i^j\in S_i$ passes the unit test $u_i$, in other words, its $C_i$ is not empty. Equation 2 is the concrete definition for $Pass@k$.

\begin{equation}
    pass@k = \frac{|\{p_i{\ |\ p}_i\in P,\left|C_i\right|>0\}|}{n}
\end{equation}

\medskip

The ultimate goal of this work is to improve the syntactical variations in correctly generated solutions (i.e., $\{C_i\}_{i=1}^n$ ). To measure the code mutation capability of $M^\theta$, we utilize the $variation@k$ metric proposed by Madani \cite{madani2023metamorphic}. Let $V_i$ denote the set of unique correct solutions from $C_i$ where $\left|V_i\right|\le\left|C_i\right|$, Then, $variation@k$ is calculated by taking the average number of distinct solutions over the total number of code instances synthesized for each problem: 

\begin{equation}
variation@k=\frac{1}{nk} \sum_{{i=1\atop\left|C_i\right|>0}}^{n}\left|V_i\right|
\end{equation}
\medskip

At first glance, this problem setup depicts a multi-objective retraining scenario. Typically, LLM-based program synthesizers are trained in an autoregressive manner with Cross-Entropy Loss as the objective function minimized during training. It is natural for one to add $variation@k$ definition also to the objective function in order to encourage more distinct code outputs with similar semantics. However, at the current state, the $variation@k$ and $pass@k$ metrics are not differentiable, which means they cannot be directly used as a loss function in gradient-based optimization methods. Also, Madani \cite{madani2023metamorphic} demonstrated that pre-trained LLMs capable of code synthesis exhibit some degree of code mutation capabilities without explicitly being trained for code mutation. Therefore, in this work, we attempt to investigate if the composition (i.e., a dataset used in code mutation training, the definition is given in Section 4) of the fine-tuning dataset could allow for learning a parameter set $\theta^\prime\neq\theta$ such that improve the code mutability of the model $M$ as measured by  $variation@k$ metric.

When a pre-trained model undergoes retraining, its problem-solving ability, measured by $pass@k$, might change. Ideally, we aim to increase or at least maintain $pass@k$ after code mutation training. However, in practice, this may not always occur. Thus, it is crucial during code mutation training to evaluate $variation@k$ alongside a metric that measures code synthesizing ability. Nonetheless, $pass@k$ does not necessarily provide sufficient information about code synthesizing ability. $Pass@k$ remains unaffected whether a code synthesizer solves a problem once or multiple times. Moreover, $pass@k$ exhibits somewhat discrete behavior. As it was mentioned, $pass@k$ is calculated as the fraction of solved problems relative to the total number of evaluation problems, regardless of variations. Each solved problem significantly increases the metric by a jump/step. This phenomenon results in a stepwise movement in $pass@k$, which obscures gradual improvements in model synthesis performance. Therefore we are going to define a new metric, $correct@k$ as
\begin{equation}
    correct@k = \frac{1}{nk}\sum_{i=1}^{n}|C_i|
\end{equation}
in order to measure the correctness of generated solutions by $M$ in tandem with $variation@k$.

The metric $correct@k$ quantifies the number of correct solutions synthesized by the model. A decrease in this metric signifies that the model is generating fewer code samples overall. Therefore increase in  $variation@k$ without a decrease in $correct@k$ is indicative of the model's ability to produce a greater variety of code while also maintaining the accuracy of these codes. This scenario is indicative of a genuine improvement in the model's code mutation capabilities
A model with a higher $variation@k$ demonstrates greater code mutation capability. However, since $variation@k$ does not include unsolved problems in its formulation, merely increasing this metric is insufficient. Specifically, if problems with initially few correct samples subsequently yield no correct samples, the $variation@k$ metric will paradoxically increase, as it does not consider the total absence of correct solutions. Therefore, one cannot claim to have achieved superior code mutation capability in M if the $variation@k$ score increases while the $correct@k$ score is decreased post fine-tuning. Therefore, effective code mutation training should ensure that
\begin{equation}
    correct@k(M, \theta^{\prime}) \geq correct@k(M , \theta)
\end{equation}
\bigskip

\section{Dataset Generation}
It is important to reiterate the fact that the goal of this work is to increase code mutation capability of a lightweight LLM   that could potentially be used by malware creators embedded in their malware binary to act as a mutation engine. During our study, we have confirmed that really large open-source LLMs such as Llama 3 with over 8 billion trainable parameters can do an acceptable job in code mutation. However, one can be certain that LLMs with over 1 billion parameters will never be embedded in metamorphic malware to act as a mutation engine due to their large disk size. 
 
To conduct the code mutation training proposed in Section 3, a suitable dataset is necessary to fine-tune a chosen pre-trained program synthesizer. Ideally, this dataset should contain variations of subroutines that are syntactically different but semantically equivalent. Currently, such a dataset specifically designed for code mutation training does not exist. To address this challenge and bootstrap our training task, we propose a framework, depicted in Figure 1, to build a dataset tailored for code mutation training. This framework employs a teacher LLM that is capable of generating instances of the training dataset. This teacher LLM will be realized using a very large LLM-based code synthesizer capable of demonstrating an acceptable degree of code mutation (as defined in Section 3) while its disk size is prohibitive to be used as an embedded mutation engine.

\begin{figure}[htbp]
\centerline{\includegraphics[width=\linewidth]{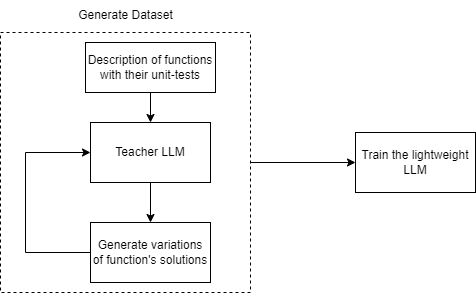}}
\caption{Framework to generate code mutation dataset}
\label{fig}
\end{figure}

The process of creating the dataset (depicted in Figure 1) begins by providing initial descriptions of the subroutine (e.g., $p_i$) to be generated to the teacher LLM; then, the teacher model, in turn, will generate the subroutine realizing the given input prompt. Finally, the generated subroutine (e.g., $s_i$) will be tested against its associated unit test (e.g., $u_i$) before the generated subroutine is added to our training dataset.

In this study, we used \textit{Top-P} sampling to generate the output. This method, proposed by Holtzman \textit{et al}. \cite{holtzman2019curious}, selects tokens for text generation based on a cumulative probability threshold. It ensures that only the most probable tokens, those whose combined probabilities exceed the threshold $p$, are considered. In generative models, when utilizing sampling strategies, the output text from models is stochastic. This stochastic nature is particularly useful in tasks like code mutation, where generating variations of code that maintain correctness can lead to new and useful modifications. Therefore, as depicted in Figure 2, each prompt is queried $k$ times against the teacher LLM in order to build a training code mutation dataset consisting of instances many of which have identical semantics.

\begin{figure}[htbp]
\centerline{\includegraphics[width=0.68\linewidth]{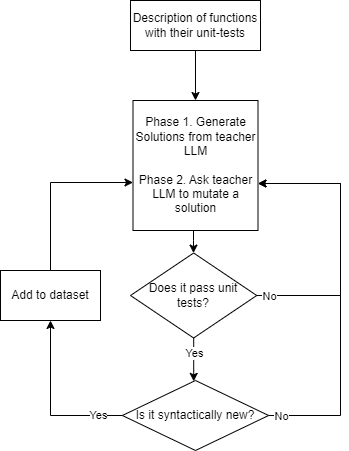}}
\caption{Iteration loop for dataset generation}
\label{fig}
\end{figure}

\begin{figure*}[htbp]
\centerline{\includegraphics[width=\linewidth]{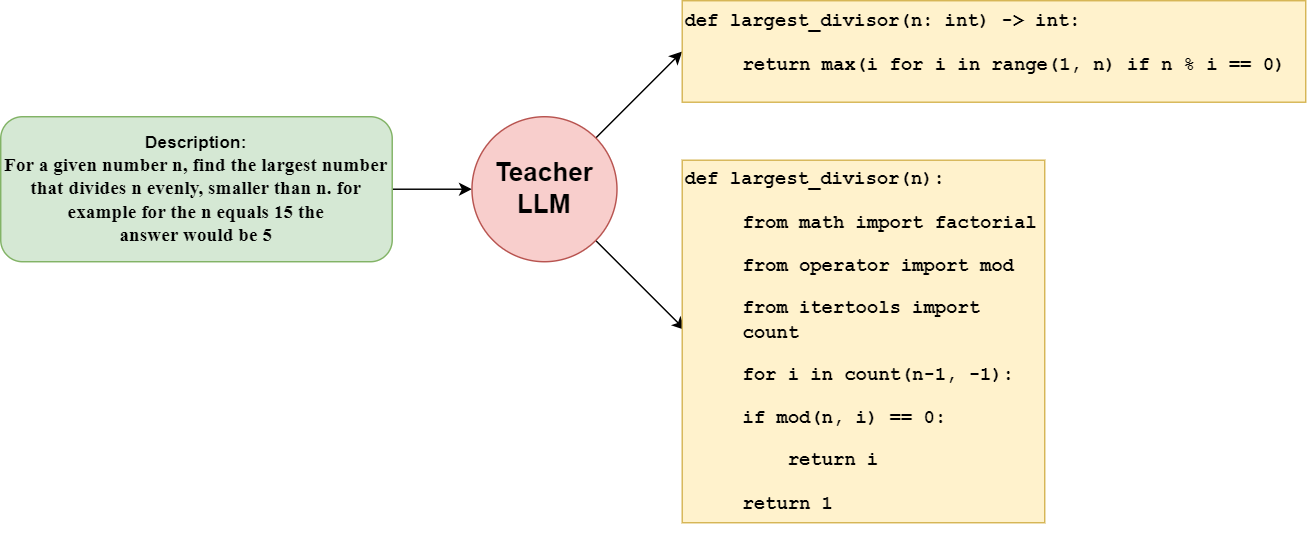}}
\caption{Training data examples}
\label{fig}
\end{figure*}

When the teacher LLM synthesizes codes for a subroutine, the output might not always be clean, meaning it often includes extra subroutine definitions that are irrelevant to the prompt description itself. These results might contain usage examples of the subroutine itself, which are redundant and could potentially require interaction with user input, leading to interruptions. Such extraneous content not only lacks utility but also presents an opportunity for code errors. In this research, we discard the extraneous content and only keep the clean subroutine definition body.

In our experiment, we used Python functions from the HumanEval dataset \cite{chen2021evaluating} as subroutines to be mutated. HumanEval is a benchmark dataset that comprises 164 Python problems, each accompanied by a description and a set of unit tests. The inclusion of necessary unit tests makes this dataset particularly suitable for our study. These problems vary in difficulty, providing an effective measure of an LLM's code generation capabilities. Since our focus is not on the code generation capabilities of the LLM and benchmarking with HumanEval but rather on enhancing the code mutation of the model, we utilize the HumanEval dataset to create different variations for the code mutation training dataset. 

We have used $llama3-instruct-8B$ \cite{dubey2024llama} as our teacher LLM. This model is a highly capable code synthesizer, making it an ideal choice for bootstrapping our training dataset. 
For the target model that we intend to fine-tune, we selected $codegen-mono-350M$, which, in comparison to llama3, has significantly fewer parameters and can be considered a lightweight LLM. Its ability to synthesize code is also considerably less than that of the teacher model. These characteristics make CodeGen-350M an excellent candidate for retraining to enhance its code mutation capabilities.

The dataset generation process starts by setting aside 15 HumanEval problems for evaluation, ensuring the model does not encounter any variations of these problems during the training phase. The rest of the prompts are input into the teacher LLM to synthesize initial solutions for the dataset. To enhance the diversity of solutions, the teacher model is tasked with writing Python functions based on the given prompts. After post-processing the generated outputs (i.e., generated solutions) and validating them for correctness and uniqueness, they are incorporated into our training dataset.

In the next step, for each problem, one of the previously written solutions is selected, and, along with the original description, it is submitted to the teacher LLM. The task for the LLM is to rewrite the function in a new syntax while preserving the semantic integrity of the function. As in the earlier phase, all synthesized codes are subjected to post-processing to remove any extraneous elements. Before being added to the dataset, each piece of code undergoes validation using the unit tests provided by the HumanEval dataset. Only those solutions that are correct and not already present in the dataset are accepted and added to the training set. This iterative process is repeated a number of times to build a growing collection of multiple unique solutions for each prompt in the HumanEval set. 

Figure 3 depicts an example of an input prompt and two unique output solutions generated by the teacher model. These sample outputs showcase the capability of the teacher LLM (and by extension, the quality of the produced dataset in general) in producing solutions using different programming approaches. By utilizing our framework we have managed to generate a mutation training dataset with 4000 distinct and correct solution variations to the initial HumanEval prompts. With the training data now prepared, the next step involves fine-tuning the lightweight LLM.

\bigskip

\section{RESULTS AND DISCUSSIONS}

In our experiment, we utilized the pre-trained model $codegen-mono-350M$ \cite{chen2021evaluating} as the target model for retraining. This model, which is built on top of the Codegen \cite{chen2021evaluating} model, has been specifically retrained on Python codes, as indicated by the `mono' term in its name. It comprises 350 million parameters, which is relatively smaller compared to state-of-the-art LLMs like llama3, which has two versions of 8 billion and 70 billion parameters. We used batches of code samples for retraining, each with a size of 8, and conducted the process over 5 epochs. The cross-entropy loss function was employed, which is commonly used in sequence generation tasks. The code mutation training was then carried out on this model using the generated dataset (described in Section 4).

After fine-tuning, to measure the model's code mutation capability, we utilize the $variation@k$ metric. As presented in Table 2, after code mutation training the average $variation@10$ has increased by approximately 15\%. This increase demonstrates the success of code mutation training and the proposed framework in dataset generation. In other words, the lightweight model, here CodeGen, showed an enhanced ability to produce a broader range of solution variations post-training.

\begin{table}[htbp]
\caption{variation@10, correct@10 and pass@10 results before and after code mutation fine-tuning}
\begin{center}
\begin{tabular}{|c|c|c|c|}
\hline

\textbf{Model} & \textbf{\textit{Variation@10}}& \textbf{\textit{Pass@10}}& \textbf{\textit{Correct@10}} \\
\hline
Codegen-mono-350M & 30 & 46 & 16.7 \\
\hline
\begin{tabular}{@{}c@{}} Codegen-mono-350M- \\ CodeMutationFine-Tuned\end{tabular}  & 44 & 32  & 18.8 \\
\hline

\end{tabular}
\label{tatheis2015note}
\end{center}
\end{table}

Figure 4 illustrates the relationship between $variation@10$ and $pass@10$ metrics before and after training, across multiple trials. As previously discussed, $variation@10$ has increased as a result of targeted training. Conversely, $pass@10$ has decreased, indicating that the model now solves fewer problems than it did prior to training. An extensive review revealed that the model tends to forget how to solve the more challenging problems, those that it could only solve once or twice out of ten attempts (in $pass@10$) before training. Additionally, the model's success with those challenging problems (i.e., prompts) was inconsistent, occasionally failing to solve them at all. In other words, before retraining and under the pass@1 metric, these problems had a near-zero chance of being solved. After undergoing code mutation training, the model exhibits even lower efficacy in solving these challenging problems, although it now solves other problems with more variations (i.e., more mutability).

\begin{figure}[htbp]
\centerline{\includegraphics[width=\linewidth]{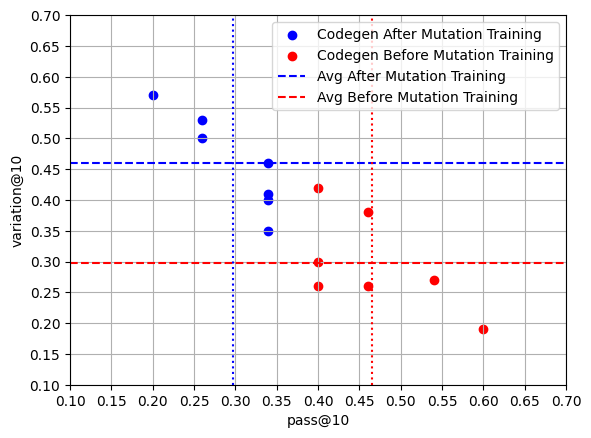}}
\caption{Comparison of variation@10 over pass@10 of Codegen before and after mutation training}
\label{fig}
\end{figure}

Another important observation is the discrete nature of the $pass@k$ metric. Since $pass@k$ is calculated as the fraction of solved problems over the total number of evaluation problems, when a problem is solved, regardless of the frequency, it significantly impacts the $pass@k$ metric. This does not reflect in $correct@k$ since it is the number of times the model successfully solves the problem. 

Figure 5 illustrates the relationship between $variation@10$ and $correct@10$ across multiple trials. The figure shows that the training not only succeeded in increasing the variations of the generated samples but also improved the average generation of correct samples for the problems. This improvement is logical, as generating more correct samples likely leads to a greater diversity of unique and distinct samples.
\begin{figure}[htbp]
\centerline{\includegraphics[width=\linewidth]{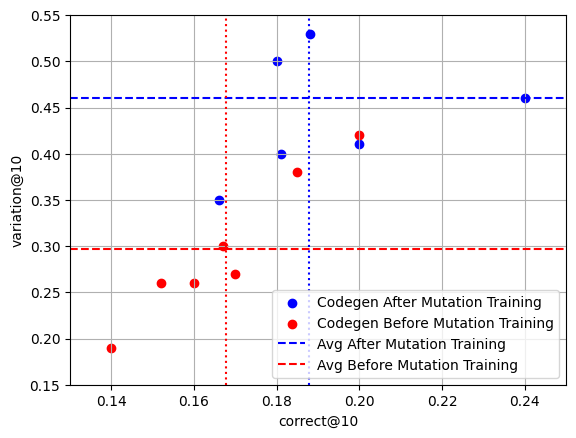}}
\caption{Comparison of variation@10 over correct@10 of Codegen before and after mutation training}
\label{fig}
\end{figure}

To examine the model's forgetfulness in solving challenging problems after training,  layer-freezing experiments were conducted. Considering the architecture of $codegen-350M$, which consists of 19 layers, different configurations were tested by $(a)$ freezing no layers, $(b)$ the first five layers, $(c)$ the first ten layers, and $(d)$ the first fifteen layers. The results of these experiments, depicted in Figure 6, indicated that freezing layers helps the model retain its ability to solve difficult problems, thereby maintaining the $pass@k$ metric. However, this approach also prevents the model from enhancing its code mutation capabilities, which is counterproductive to the objectives of this study.

\begin{figure}[htbp]
\centerline{\includegraphics[width=\linewidth]{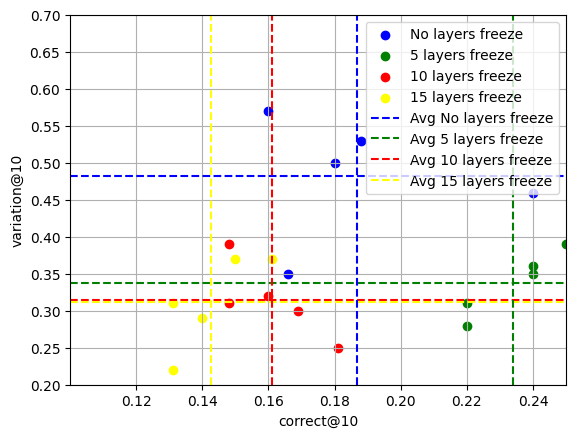}}
\caption{Comparison of variation@10 over correct@10 for different layer freezing conditions}
\label{fig}
\end{figure}

Furthermore, comparing the $correct@10$ metric across different layer freezing configurations reveals that freezing the first five layers yields the highest number of correct samples. This indicates that the code mutation training led to a model capable of generating more correct samples. However, since $variation@10$ did not increase significantly, it suggests that most of the correct solutions are similar and not unique compared to each other, thus making the mutation training not a strong success in achieving code mutation. In contrast, the model with no layer freezing exhibited the highest code mutation capability, indicating that training all layers is necessary for generating both correct and varied code samples.

\bigskip
\section{Concolusion AND Future Works}
In conclusion, in this study, we proposed a concrete definition for code mutation training of LLM-based code mutation engines. We outline how LLM-based synthesizers can be specifically fine-tuned to generate more varied code (i.e., mutate code) and how to evaluate their mutation capabilities. Additionally, we proposed a framework and generated the first publicly available dataset for code mutation training using a teacher LLM. We demonstrated that by leveraging the expertise of the teacher LLM to build the dataset, the lightweight LLM has been effectively trained to synthesize code with increased variations and correctness. Additionally, we carried out several layer-freezing experiments to analyze their impacts.

For this study, we targeted a lightweight LLM, which, although less complex than state-of-the-art LLMs, could still enhance its code mutation capabilities. While the lightweight model still requires considerable infrastructure to generate results, it confirms that it is feasible to fine-tune a code-synthesizer LLM for code mutation. Additionally, recent studies aimed at compressing LLMs support the notion of seamlessly integrating an LLM-based mutation engine into the core of malware to alter its form and syntax in the future.

The use of LLMs in developing (malicious) code mutation engines is truly a new frontier in Software Engineering and Cybersecurity with greatly many open questions. As one of the future research directions, we plan to expand the dataset generation framework by incorporating a larger set of problems, aiming to create a more extensive training dataset. A larger and more diverse dataset is expected to enhance the performance of the trained models and further increase their $variation@k$ score. Also, it is essential to explore methods for conducting code mutation training without any (or much less)  forgetting and therefore no loss in $pass@k$ score. The goal will be to enhance the model's code mutation capabilities while preserving its proficiency in code synthetization, ensuring no compromise in overall generation performance.

Moreover, different approaches for retraining should be explored. Employing other modeling approaches, such as sequence-to-sequence (seq2seq) for fine-tuning, could yield better results and achieve higher $variation@k$ in the model.

In addition, conducting similar experiments with various teacher and target LLMs could yield further insights, particularly regarding the trainability of code mutation across different pre-trained code synthesizer models. Moreover, utilizing different teacher LLMs may facilitate the production of larger and cleaner datasets. Additionally, replicating this research with diverse lightweight LLMs and comparing the outcomes can enhance our understanding of code mutation capabilities. 

Extending these experiments to encompass other programming languages would broaden the applicability and deepen the understanding of the models' capabilities in diverse coding environments. Furthermore, now that we have demonstrated the feasibility of code mutation training on lightweight LLMs, future efforts should focus on strengthening defenses against potential cyber threats that could emerge. Developing robust countermeasures will be crucial for safeguarding against such evolving risk.

\bibliographystyle{IEEEtran}
\bibliography{references}

\end{document}